\documentclass[a4paper,11pt]{article}
\usepackage{pos}
\usepackage{sidecap}
\sidecaptionvpos{figure}{t}
\usepackage{color}
 
\usepackage{setspace}

\title{Simulation and sensitivities for a phased IceCube-Gen2 deployment}
 \ShortTitle{Simulation and sensitivities of Gen2}
 
\author{The IceCube-Gen2 Collaboration \\{\normalsize \normalfont(a complete list of authors can be found at the end of the proceedings)}}

\emailAdd{baclark@msu.edu}
\emailAdd{hallid15@msu.edu}

\abstract{The IceCube Neutrino Observatory opened the window on high-energy neutrino astronomy by confirming the existence of PeV astrophysical neutrinos and identifying the first compelling astrophysical neutrino source in the blazar TXS0506+056. Planning is underway to build an enlarged detector, IceCube-Gen2, which will extend measurements to higher energies, increase the rate of observed cosmic neutrinos and provide improved prospects for detecting fainter sources. IceCube-Gen2 is planned to have an extended in-ice optical array, a radio array at shallower depths for detecting ultra-high-energy (>100 PeV) neutrinos, and a surface component studying cosmic rays. In this contribution, we will discuss the simulation of the in-ice optical component of the baseline design of the IceCube-Gen2 detector, which foresees the deployment of an additional $\sim$\,120 new detection strings to the existing 86 in IceCube over $\sim$\,7 Antarctic summer seasons. Motivated by the phased construction plan for IceCube-Gen2, we discuss how the reconstruction capabilities and sensitivities of the instrument are expected to progress throughout its deployment.

\vspace{4mm}
{\bfseries Corresponding authors:}
Brian A. Clark$^{1*}$, Robert Halliday$^{1}$\\
{$^{1}$ \itshape Michigan State University, East Lansing, MI, United States}\\
[4mm]
$^*$ Presenter

\FullConference{37$^{\rm{th}}$ International Cosmic Ray Conference (ICRC 2021)\\
		July 12th -- 23rd, 2021\\
		Online -- Berlin, Germany}

}

\begin{document}
\maketitle

\section{Introduction}

Neutrinos are unique messengers to the distant and  high-energy reaches of the Universe. Unlike cosmic rays and gamma rays, neutrinos are chargeless and weakly interacting particles, capable of travelling in straight lines from their birthplace in astrophysical accelerators. The IceCube observatory opened the window on high-energy neutrino astronomy by detecting an astrophysical neutrino flux in 2013~\cite{Aartsen:2013jdh}. IceCube has characterized the flux of astrophysical neutrinos, concluding it is consistent with an isotropic arrival direction distribution. A fit to the energy ($E$) distribution of the form a single single power law $E^{-\gamma}$ yields a spectral index $2<\gamma<3$. It also appears  the three neutrino flavors contribute roughly equally to the total flux~\cite{Abbasi:2020jmh,Aartsen:2020aqd, Aartsen:2018vez, Stettner:2019tok}. In addition to measuring this diffuse flux of neutrinos, IceCube has also searched for the sources of neutrinos, identifying the first compelling evidence of a neutrino source in the blazar TXS0506+056~\cite{IceCube:2018dnn, IceCube:2018cha}. The IceCube detector itself is an array of 5,160 photomultiplier tubes deployed instrumenting a cubic kilometer of clear glacial ice near the geographic South Pole, and is designed to detect the Cherenkov light emitted by charged particles produced in neutrino interactions in the ice.

To discover ultra-high-energy neutrinos above 10\,PeV, better characterize the flux of astrophysical neutrinos with a larger sample size, and identify more neutrino sources, a new, larger detector is needed. To meet this need, the IceCube-Gen2 detector is under development~\cite{Aartsen:2020fgd}. IceCube-Gen2 will feature an extended in-ice optical array, which is the focus of this proceeding, as well as a new shallow-radio array for the detection of ultra-high energy neutrinos, and a surface array for studying cosmic rays and for providing a veto to downgoing atmospheric neutrinos. IceCube-Gen2 will play an important and complimentary role to other next-generation neutrinos telescopes---such as KM3NeT~\cite{Adrian-Martinez:2016fdl}, Baikal-GVD~\cite{1997APh.....7..263B}, P-ONE~\cite{Agostini:2020aar}, etc.---in characterizing the astrophysical neutrino flux and searching for sources of neutrinos.

The construction of the IceCube-Gen2 detector is expected to take approximately seven years. 
Similarly to IceCube, during the construction phase the  Gen2 array can already deliver a compelling scientific program. We discuss how the science capabilities evolve with the partially constructed array. In Sec.~\ref{sec:instrument} we describe in the enlarged optical component of the IceCube-Gen2 facility. In Sec.~\ref{sec:sim_and_fom} we describe the simulation and performance of the detector, and in Sec.~\ref{sec:sensitivity} we describe the resulting expected sensitivity. Finally, in Sec.~\ref{sec:conclusion} we review our results and discuss the outlook going forward.

\section{The Gen2-Optical Instrument}
\label{sec:instrument}

\begin{SCfigure}
    \centering
    \includegraphics[width=0.4 \textwidth]{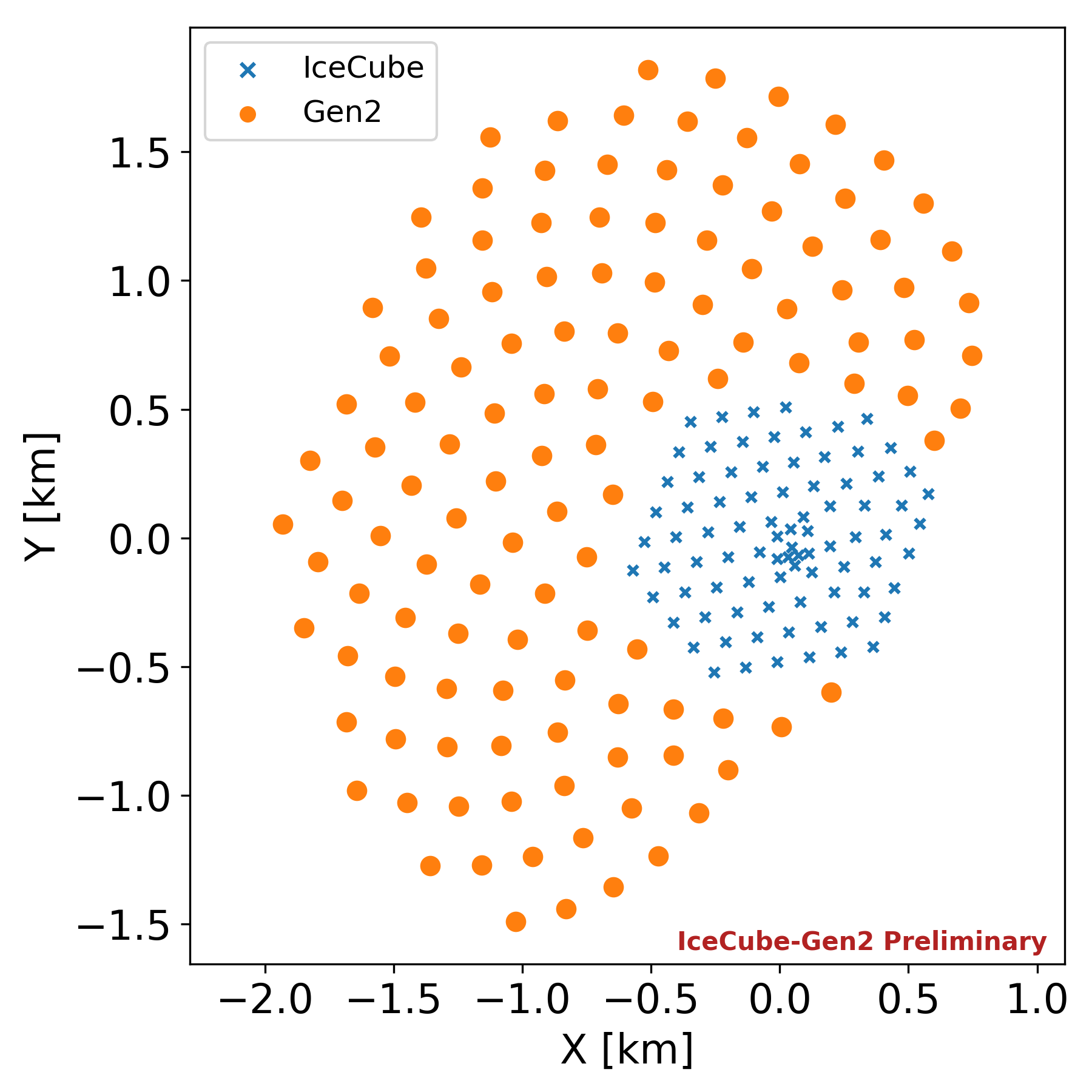}
    \caption{A top-down view of the IceCube-Gen2 optical detector, with the original 86 IceCube strings marked with crosses, and the 120 new Gen2 strings marked with circles.}
    \label{fig:gen2_layout}
\end{SCfigure}

In this section, we describe the enlarged optical array, which is shown in a "top-down" view in Fig.~\ref{fig:gen2_layout}. The array consists of 120 new strings of optical modules (OMs) deployed in a "sunflower" pattern around the original 86 IceCube strings. The total volume of the Gen2-Optical detector is $\sim$8\,km$^3$, about an order of magnitude larger than the IceCube detector. As can be seen in Fig.~\ref{fig:gen2_layout}, the lateral spacing between strings has been roughly doubled, from 125\,m in IceCube to 240\,m. The optimization of the string spacing is the topic of another proceeding at this conference~\cite{Omeliukh:2021icrc_opt}. The vertical spacing between OMs is 17\,m, the same as in IceCube;  however, the OMs are distributed across a slightly larger range of depths, from 1340-2700\,m in depth, instead of 1446-2451\,m.

\begin{figure}[htp]
    \centering
    \includegraphics[width=0.49 \textwidth]{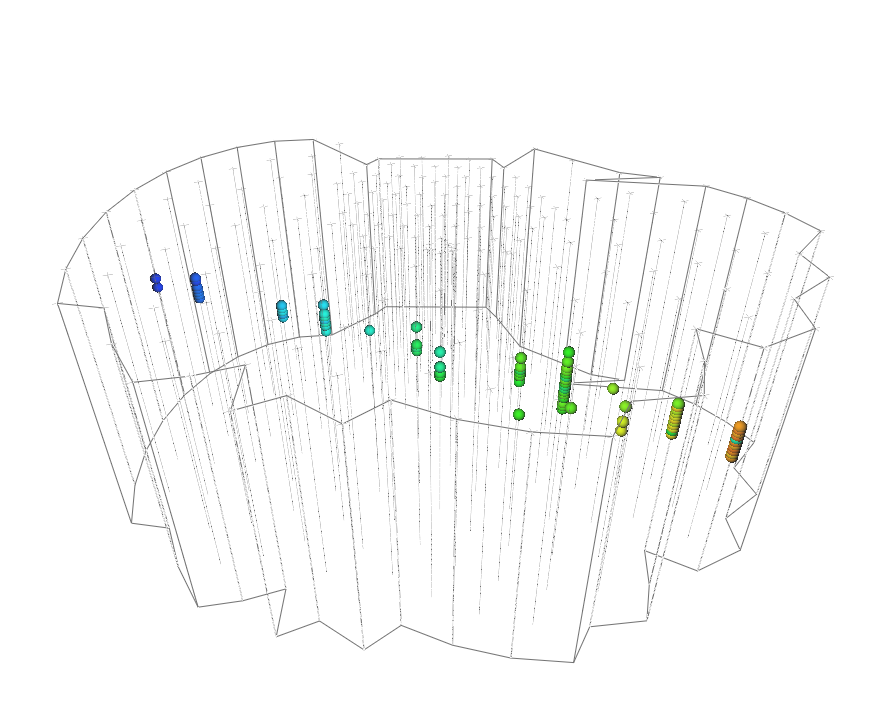}
    \includegraphics[width=0.49 \textwidth]{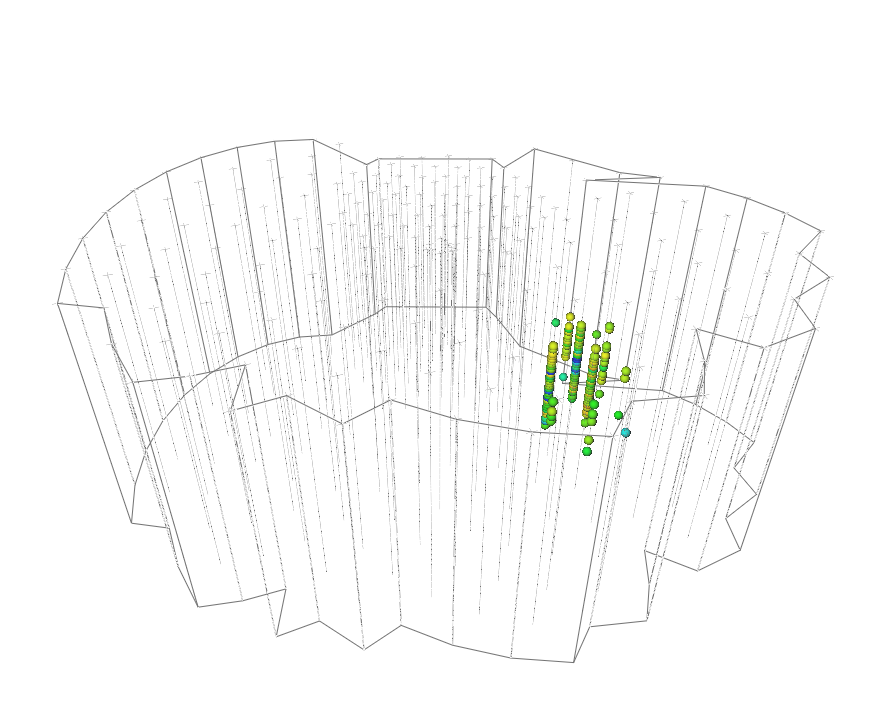}
    \caption{A view of the two main ``morphologies" of neutrino interactions expected in the optical component of IceCube-Gen2: ``tracks" (left) ``cascades" (right). The size of a module is proportional to the quantity of the light recorded, while the color indicates the relative arrival time of the light (with red indicating earlier arrival times, and blue indicating later arrival times).}
    \label{fig:events}
\end{figure}

Like in IceCube, neutrinos are expected in IceCube-Gen2 in two main light deposition patterns, or "morphologies;" both are visible in Fig.~\ref{fig:events}. The first, called ``tracks," primarily arise from charged-current interactions of muon neutrinos, giving rise to long-lived daughter muons which cross the detector. The second, called "cascades," are roughly spherical depositions of light arising from neutral-current interactions of all flavors and charged-current interactions of electron and tau neutrinos. At sufficiently high energies ($\sim 1$\,PeV) tau neutrinos can also produce additional  interesting morphologies; for example, the daughter tau can travel macroscopic distance from the primary vertex before decaying, resulting in a secondary deposition of light in a signature termed a ``double bang". At EeV energies, the daughter taus become sufficiently long lived to appear as tracks.

For these proceedings, we focus on the performance of the detector to a sub-class of events known as ``throughgoing-tracks," as this class of events historically provides the best sensitivity in searches for neutrino sources~\cite{Aartsen:2019fau}. Such events are largely produced by muons that cross the detector. The muons can either arise from astrophysical and atmospheric neutrinos interacting outside of the detector volume in both hemispheres, or from atmospheric muons in the southern hemisphere.

\section{Simulation and Performance}
\label{sec:sim_and_fom}
In this section, we describe the process of simulating the optical component of IceCube-Gen2 from particle generation, through photon propogation, detector simulation, processing and reconstruction. From this simulation we extract relevant performance metrics, specifically the effective area and pointing resolution.

\subsection{Simulation and Reconstruction}
The first step of our simulation is to specify the primary particle that will be generated. Since we focus on throughgoing tracks, we simulate primaries as single high-energy muons. These muons are injected with the MUONGUN~\cite{JVSThesis} simulation package, and then further propagated with PROPOSAL~\cite{Koehne:2013gpa}, which estimates the energy depositions of the muons from continuous and stochastic processes such as ionization and bremsstrahlung. Muons are simulated as arriving isotropically from all directions on the sky and are distributed in energy between 3\,TeV and 100\,PeV according to a power law $E^{-1.4}$ spectrum; the hard spectral index ensures sufficient statistics to characterize the detector performance at high energies.

\begin{SCfigure}
    \centering
    \caption{The acceptance to photons as a function of wavelength for two different types of optical modules used in IceCube: the standard IceCube OM, and a high quantum-efficiency variant used in DeepCore.}
    \includegraphics[width=0.5 \textwidth]{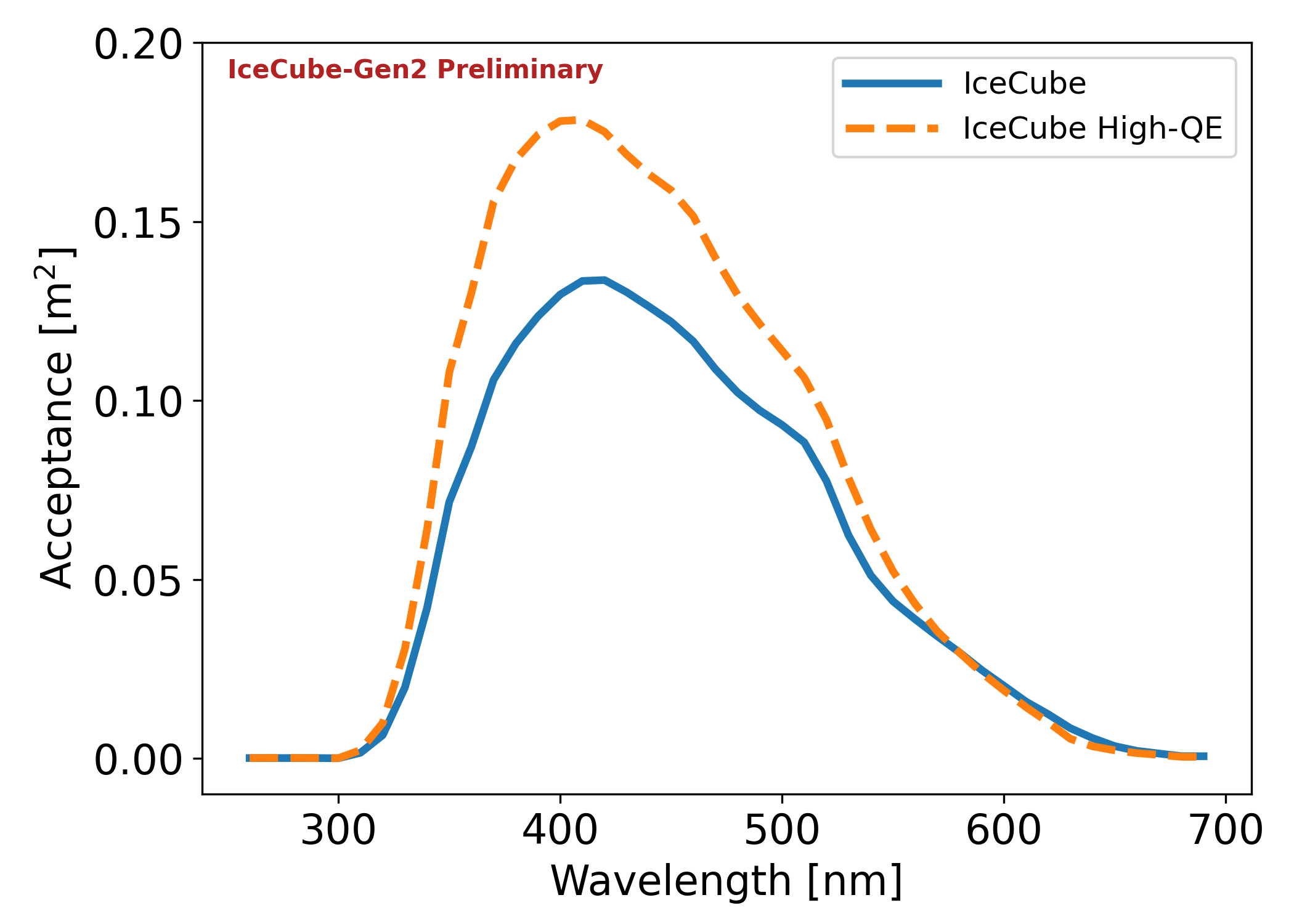}
    \label{fig:wlen_acceptance}
\end{SCfigure}

After energy depositions are specified, photons resulting from the interactions are propagated with either the CLSim or PPC package~\cite{9041727}, taking into account scattering and attenuation from the ice. The intensity of Cherenkov photons goes as $1/\lambda^2$ (where $\lambda$ is the wavelength of light), and so the spectrum is peaked in the ultra-violet; however, the response of the OMs is not, as can be seen in Fig.~\ref{fig:wlen_acceptance}. As such, to be computational efficient, photons are drawn from the Cherenkov spectrum weighted by the wavelength acceptance of the OMs. Because the ray tracing does not know ahead of time which type of module a photon will encounter, photons are generated according to an envelope which encompasses the acceptance of all modules in the simulation. After arriving at a specific module, the photons are downsampled according to the acceptance and geometry of that specific module.

After the photons arrive at the module, they must be converted into measured photoelectrons. For OMs in the original 86 IceCube strings, we apply the full IceCube simulation and pulse unfolding process: the photons are convolved with the response of the PMT and electronics, resulting in waveforms; those waveforms are then deconvolved back into a reconstructed pulse series. OMs for IceCube-Gen2 are in the conceptual design phase, hence no full PMT or electronics simulation is yet available. Instead, the photoelectrons are binned in time (mimicking the effects of a digitizer) and converted directly into reconstructed pulses.

Finally, a likelihood based reconstruction is applied to estimate the direction and energy of the muon. The method compares the observed number of photons to that expected for a hypothetical incident muon with incoming direction $\theta, \phi$ and energy $E$, and minimizes the difference in a likelihood to identify a best-fit hypothesis~\cite{Aartsen:2013vja}.

\subsection{Performance}
\label{subsec:perf}
For evaluating detector performance, we focus on two metrics that drive the sensitivity of analyses utilizing throughgoing tracks, e.g. astrophysical neutrino source searches. These are the effective area and the angular resolution. We begin by applying cuts to identify well-reconstructed tracks. This is done by imposing requirements on the reconstructions, such as requiring a minimum number of hit OMs, a minimum reconstructed track length, etc. The energy ($E$) and zenith ($\theta$) dependent efficiency for passing these cuts defines the \textit{selection efficiency}: $\eta(E, \theta)$.

To calculate the muon effective area in a given energy and zenith bin, we take the product of the projected geometric area of the detector $A_{\rm{geo}}(\theta)$ and the selection efficiency $\eta(E, \theta)$. $A_{\rm{geo}}(\theta)$ is found by placing a convex hull around the strings, defining a solid. To calculate the angular resolution, we apply the event selection, and then calculate the opening angle, or the  difference between the reconstructed and true direction $\Delta \Psi$. The distribution of $\Delta \Psi$ defines the \textit{point spread function} of the detector, which is parameterized as a function of energy and zenith using the ``Moffat/King" function, as is done in e.g. the \textit{Fermi} gamma ray telescope~\cite{1969A&A.....3..455M, 1962AJ.....67..471K, FermiPSF}. The full parameterization $\Delta \Psi (E, \theta)$ is needed for sensitivity studies in Sec.~\ref{sec:sensitivity}. We take the median opening angle, $\Delta \Psi _{\rm{med}}$, as the angular resolution. 

\begin{figure}[htp]
    \centering
    \includegraphics[width=\textwidth]{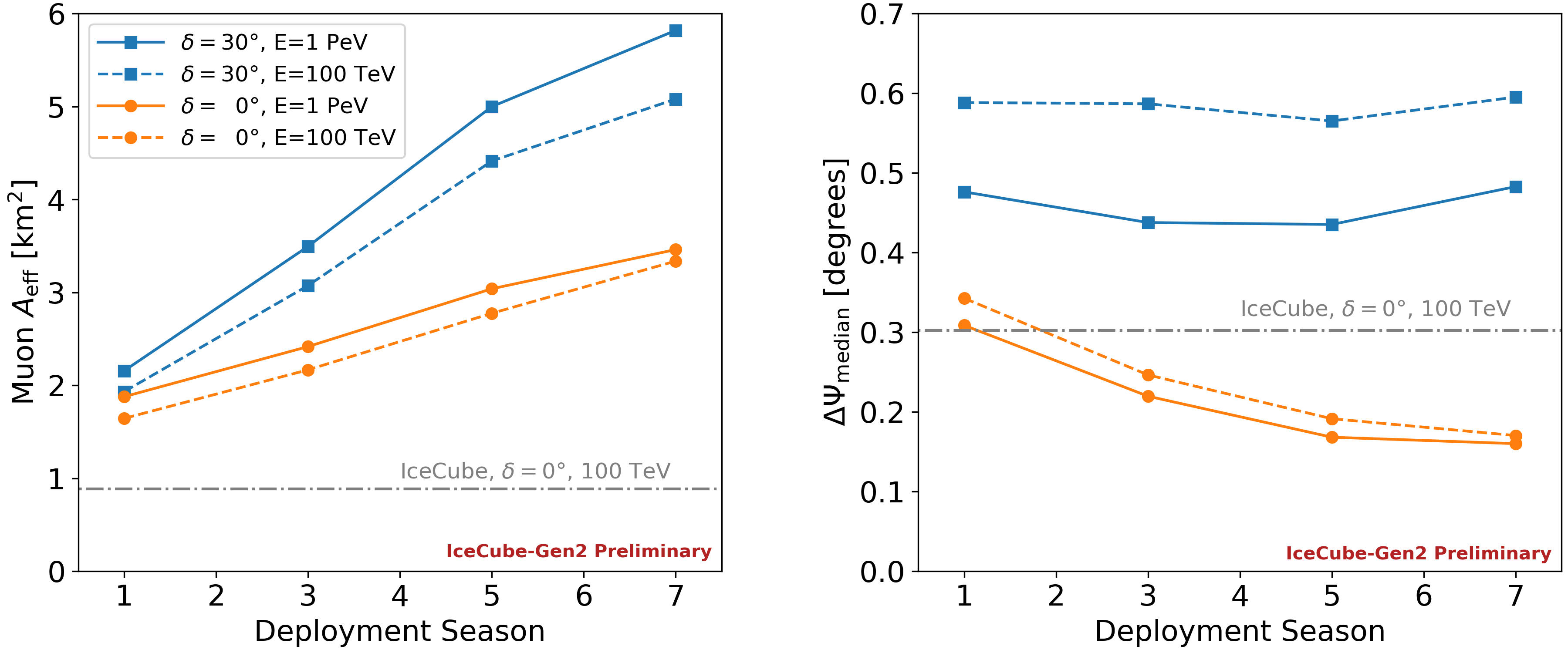}
    \caption{Plots of two metrics of the performance of the detector performance: the muon effective area (left) and the angular resolution (right) as a function of time for several choices of muon energy and declination.}
    \label{fig:performance}
\end{figure}

The two metrics, muon effective area and angular resolution, are plotted in Fig.~\ref{fig:performance}. They are plotted as a function of deployment season, and both are shown for two different energies and two different declinations; for context, we also show the performance of the IceCube detector at one representative choice of declination and energy. We assume a schedule for deployment where an equal number of strings are deployed each year. As can be seen, even by the middle of the scheduled deployment process, the total aperture for horizontal events, which provide the best reconstruction performance, will have increased by more than a factor of two, and the angular resolution will have improved by $\sim$50\%. To be conservative, these metrics do not factor in the optical modules enhanced directional reconstruction capabilities from having directional light sensing capabilities through multiple PMTs, a major improvement over IceCube DOMs.

\section{Sensitivity}
\label{sec:sensitivity}

In assessing the sensitivity of the detector, we continue to focus our science goals on the search for sources of astrophysical neutrinos. We quantify this in two different ways: a) the time-integrated sensitivity of the detector to a steady point-source of neutrinos (subsection \ref{subsec:steady}) and b) the sensitivity of the detector to a flaring source of neutrinos (subsection \ref{subsec:flare}). For both, we take the background to the neutrino search as the sum of the conventional (Honda 2006 \cite{Honda:2006qj}) and prompt (Enberg~\cite{Enberg:2008te}) atmospheric neutrino fluxes with knee reweighting according to Gaisser H3a~\cite{Gaisser:2012zz}.

\subsection{Steady Sources}
\label{subsec:steady}
\begin{SCfigure}
    \centering
    \caption{One-year time-integrated discovery potentials for a steady point source of neutrinos with a unbroken power law spectrum $E^{-2}$. Each line indicates a different year of IceCube-Gen2 construction. Also shown for reference is the 1-year discovery potential of IceCube. We emphasize that one year time-integrated discovery potentials are only shown for purpose of comparison, as the achieved discovery potential in each season will be significantly better if it is combined with the previous years data.}
    \includegraphics[width=0.60 \textwidth]{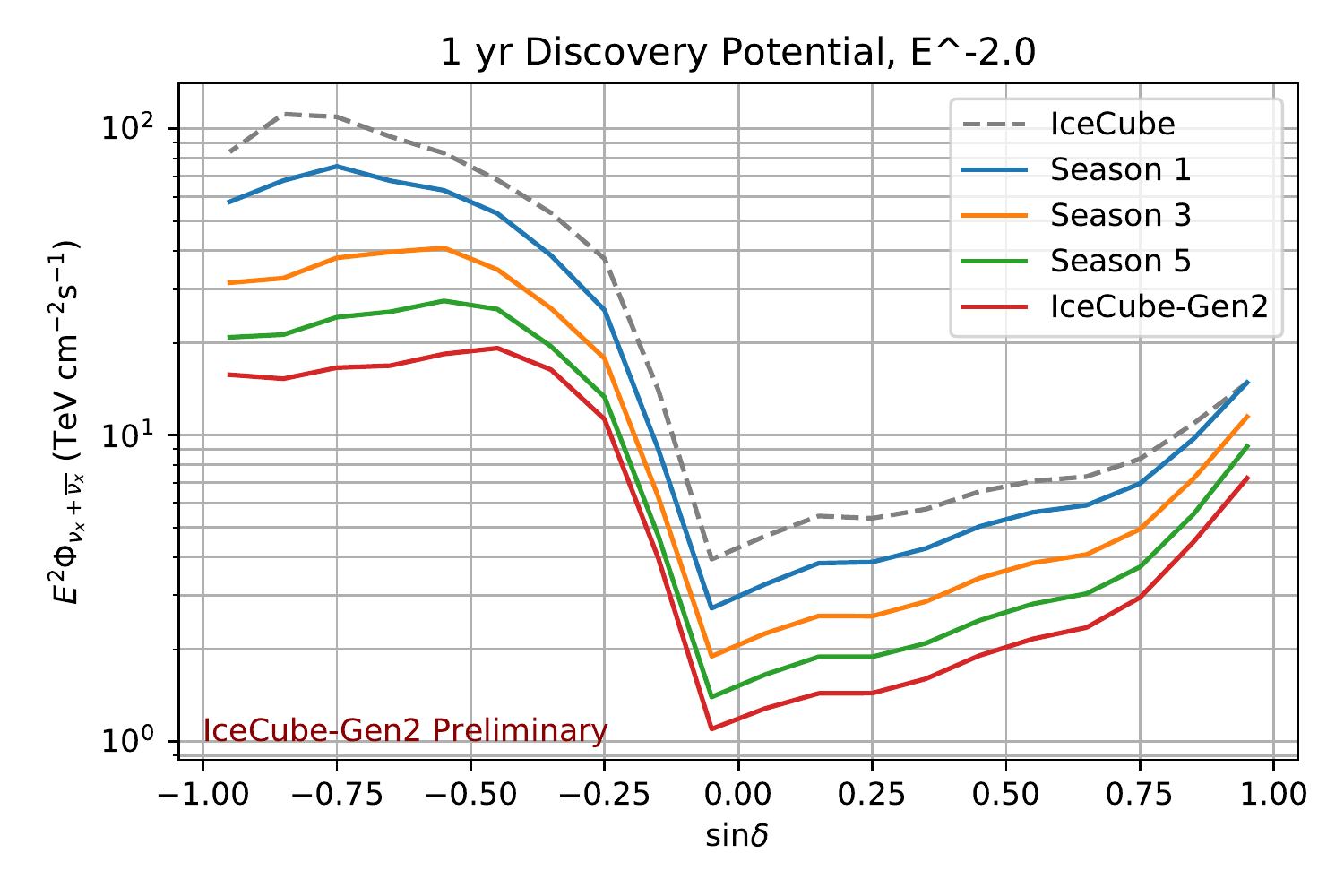}
    \label{fig:steady}
\end{SCfigure}

In modeling the sensitivity of the detector to steady point sources, we calculate the discovery potential of a mock time-integrated search. We define the discovery potential as the minimum value required of a parameter of the astrophysical neutrino flux (typically the normalization), such that an experimental search would exclude the parameter as being zero (the null hypothesis) at the $5\sigma$ level. For purposes of comparing the capabilities as a function of time, we calculate the discovery potential assuming a time integration of one year. The resulting discovery potential for IceCube-Gen2 is shown in Fig.~\ref{fig:steady}; we also show the discovery potential of the existing IceCube detector for reference. We emphasize that one year time-integrated discovery potentials are only shown for purpose of comparison, as the achieved discovery potential in each season will be significantly better if it is combined with the previous years data. The discovery potential is shown for the normalization on the flux of neutrinos, assuming an unbroken power law with a spectral index of -2. 

As in IceCube, the sensitivity of the detector is better for sources located in the Northern Sky ($\sin\delta>0$), where the Earth serves as a shield to the flux of atmospheric muons. IceCube-Gen2 also improves on the performance of IceCube near the Celestial South Pole, where an enlarged surface veto enables rejection of vertically downgoing atmospheric air showers over a broader range in zeniths. This can be seen as a flattening of the discovery potential (relative to IceCube) at $\sin\delta<-0.5$ in Fig.~\ref{fig:steady}, and is discussed further elsewhere~\cite{Aartsen:2020fgd}. As the deployment seasons progress, the discovery potential lowers to see approximately $\sim$4 times fainter sources. The sensitivity will be further improved by the  multi-PMT IceCube-Gen2 optical modules, as previously mentioned in \ref{subsec:perf}, which we have not included in these estimates.

\subsection{Flaring Sources}
\label{subsec:flare}
\begin{SCfigure}
    \centering
    \caption{Significance as a function of flare duration for various stages of IceCube-Gen2 construction including trials correction. Backgrounds here are calculated from the previously mentioned atmospheric and prompt fluxes, summed with an additional diffuse $E^{-2}$ astrophysical background component. Vertical bar in the center marks the measured flare duration in IceCube. IceCube significance included for reference.}
    \includegraphics[width=0.60 \textwidth]{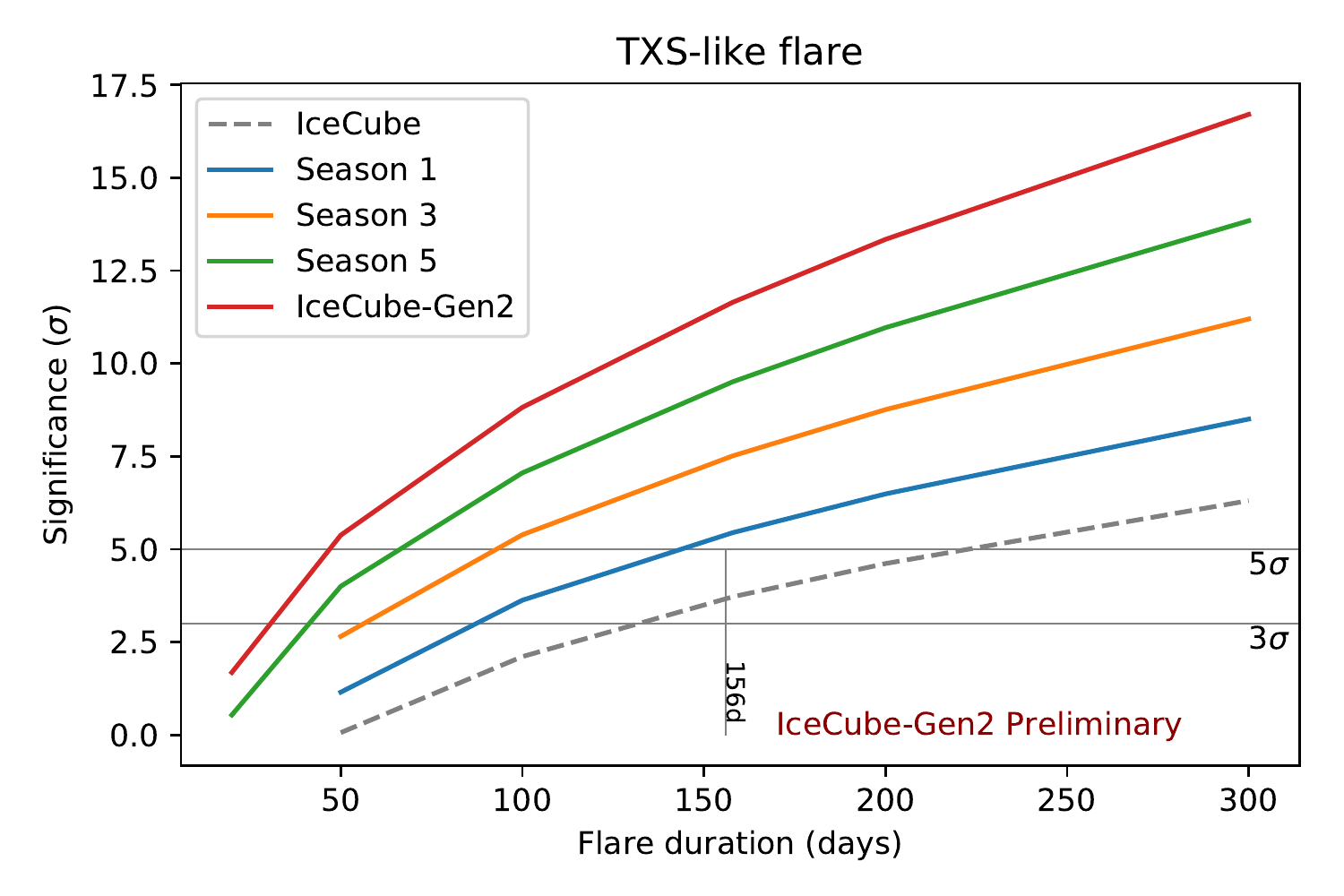}
    \label{fig:flaring}
\end{SCfigure}

To calculate the sensitivity of the detector to a flaring source of neutrinos, we consider the detection significance of a flare in IceCube-Gen2 as a function of flare duration. We take as a template the fluence of the 2014-15 flare of neutrinos from TXS0506+056, which had a measured spectral index of $\gamma=2.2
\pm 0.2$. We vary the flare duration, and calculate the significance with which it would be recorded in the various seasons of IceCube-Gen2 deployment. This "time-to-significance" can be seen in Fig.~\ref{fig:flaring}, where each line indicates a different year of the IceCube-Gen2 construction process; for reference, the bottom line shows IceCube. For a TXS-like flare, which lasted 156 days in the assumption of a box-shaped time window for the emissions, the flare could have been detected in IceCube-Gen2 at greater than $5\sigma$ significance after just the first or second season of construction (given the deployment scheduled assumed in this proceedings, where an equal number of strings is deployed each season). Importantly, this detection could be made using the neutrino search only, without any multimessenger gamma-ray counterparts, suggesting that IceCube-Gen2 will be sensitive to so called "hidden sources" which might live in gamma-ray opaque environments.

\section{Conclusion and Outlook}
\label{sec:conclusion}

In this proceeding, we have described the IceCube-Gen2 optical instrument, and the simulation process to determine sensitivities. Motivated by the $\sim 7$ year deployment schedule, we determined the performance and sensitivity of the detector over time. For the performance, we focused on throughgoing tracks, which provide the best discovery potential for astrophysical sources. We have evaluated the effective area and angular resolution of such events, finding an approximately four times greater effective area and an approximately two times improved angular resolution. Finally, we show the projected sensitivity of the detector to both steady and flaring sources of neutrinos, showing that in the time that IceCube observed the TXS flare at 3.5$\sigma$, the full IceCube-Gen2 detector would have seen it at more than 10$\sigma$. These performance characteristics show that even during construction, IceCube-Gen2 will have a rapidly increasing discovery potential for astrophysical neutrino sources and will be essential for multi-messenger science.

\bibliographystyle{ICRC}
\setlength{\bibsep}{5pt}
\bibliography{references}

\providecommand{\href}[2]{#2}\begingroup\raggedright\setstretch{0.01}\begin{thebibliography}{10}

\bibitem{Aartsen:2013jdh}
{\bfseries IceCube} Collaboration, M.~G. Aartsen {\em et~al.}
  \href{http://dx.doi.org/10.1126/science.1242856}{{\em Science} {\bfseries
  342} (2013) 1242856}.

\bibitem{Abbasi:2020jmh}
{\bfseries IceCube} Collaboration, R.~Abbasi {\em et~al.}
  \href{http://dx.doi.org/10.1103/PhysRevD.104.022002}{{\em Phys. Rev. D}
  {\bfseries 104} (Jul, 2021) 022002}.

\bibitem{Aartsen:2020aqd}
{\bfseries IceCube} Collaboration, M.~G. Aartsen {\em et~al.}
  \href{http://dx.doi.org/10.1103/PhysRevLett.125.121104}{{\em Phys. Rev.
  Lett.} {\bfseries 125} no.~12, (2020) 121104}.

\bibitem{Aartsen:2018vez}
{\bfseries IceCube} Collaboration, M.~G. Aartsen {\em et~al.}
  \href{http://dx.doi.org/10.1103/PhysRevD.99.032004}{{\em Phys. Rev. D}
  {\bfseries 99} no.~3, (2019) 032004}.

\bibitem{Stettner:2019tok}
{\bfseries IceCube} Collaboration, J.~Stettner
  \href{http://dx.doi.org/10.22323/1.358.1017}{{\em PoS} {\bfseries ICRC2019}
  (2020) 1017}.

\bibitem{IceCube:2018dnn}
{\bfseries IceCube, Fermi-LAT, MAGIC, AGILE, ASAS-SN, HAWC, H.E.S.S., INTEGRAL,
  Kanata, Kiso, Kapteyn, Liverpool Telescope, Subaru, Swift NuSTAR, VERITAS,
  VLA/17B-403} Collaboration, M.~G. Aartsen {\em et~al.}
  \href{http://dx.doi.org/10.1126/science.aat1378}{{\em Science} {\bfseries
  361} no.~6398, (2018) eaat1378}.

\bibitem{IceCube:2018cha}
{\bfseries IceCube} Collaboration, M.~G. Aartsen {\em et~al.}
  \href{http://dx.doi.org/10.1126/science.aat2890}{{\em Science} {\bfseries
  361} no.~6398, (2018) 147--151}.

\bibitem{Aartsen:2020fgd}
{\bfseries IceCube-Gen2} Collaboration, M.~G. Aartsen {\em et~al.}
  \href{http://dx.doi.org/10.1088/1361-6471/abbd48}{{\em J. Phys. G} {\bfseries
  48} no.~6, (2021) 060501}.

\bibitem{Adrian-Martinez:2016fdl}
{\bfseries KM3Net} Collaboration, S.~Adrian-Martinez {\em et~al.}
  \href{http://dx.doi.org/10.1088/0954-3899/43/8/084001}{{\em J. Phys. G}
  {\bfseries 43} no.~8, (2016) 084001}.

\bibitem{1997APh.....7..263B}
I.~A. {Belolaptikov} {\em et~al.}
  \href{http://dx.doi.org/10.1016/S0927-6505(97)00022-4}{{\em Astroparticle
  Physics} {\bfseries 7} no.~3, (Aug., 1997) 263--282}.

\bibitem{Agostini:2020aar}
{\bfseries P-ONE} Collaboration, M.~Agostini {\em et~al.}
  \href{http://dx.doi.org/10.1038/s41550-020-1182-4}{{\em Nature Astron.}
  {\bfseries 4} no.~10, (2020) 913--915}.

\bibitem{Omeliukh:2021icrc_opt}
{\bfseries IceCube} Collaboration, A.~Omeliukh
  \href{http://dx.doi.org/10.22323/1.395.1184}{{\em PoS} {\bfseries ICRC2021}
  (these proceedings) 1184}.

\bibitem{Aartsen:2019fau}
{\bfseries IceCube} Collaboration, M.~G. Aartsen {\em et~al.}
  \href{http://dx.doi.org/10.1103/PhysRevLett.124.051103}{{\em Phys. Rev.
  Lett.} {\bfseries 124} no.~5, (2020) 051103}.

\bibitem{JVSThesis}
J.~van Santen, {\em Neutrino Interactions in IceCube above 1 TeV}.
\newblock PhD thesis, University of Wisconsin-Madison, 2014.

\bibitem{Koehne:2013gpa}
J.~H. Koehne {\em et~al.}
  \href{http://dx.doi.org/10.1016/j.cpc.2013.04.001}{{\em Comput. Phys.
  Commun.} {\bfseries 184} (2013) 2070--2090}.

\bibitem{9041727}
D.~Chirkin {\em et~al.},
  \href{http://dx.doi.org/10.1109/eScience.2019.00050}{``Photon propagation
  using gpus by the icecube neutrino observatory,''} in {\em 2019 15th
  International Conference on eScience (eScience)}, pp.~388--393.
\newblock 2019.

\bibitem{Aartsen:2013vja}
{\bfseries IceCube} Collaboration, M.~G. Aartsen {\em et~al.}
  \href{http://dx.doi.org/10.1088/1748-0221/9/03/P03009}{{\em JINST} {\bfseries
  9} (2014) P03009}.

\bibitem{1969A&A.....3..455M}
A.~F.~J. {Moffat} {\em Astronomy and Astrophysics} {\bfseries 3} (Dec., 1969)
  455.

\bibitem{1962AJ.....67..471K}
I.~{King} \href{http://dx.doi.org/10.1086/108756}{{\em Astronomical Journal}
  {\bfseries 67} (Oct., 1962) 471}.

\bibitem{FermiPSF}
``The point spread function.''
  \url{https://fermi.gsfc.nasa.gov/ssc/data/analysis/documentation/Cicerone/Cicerone_LAT_IRFs/IRF_PSF.html},
  2018.
\newblock Accessed: 2021-05-28.

\bibitem{Honda:2006qj}
M.~Honda {\em et~al.} \href{http://dx.doi.org/10.1103/PhysRevD.75.043006}{{\em
  Phys. Rev. D} {\bfseries 75} (2007) 043006}.

\bibitem{Enberg:2008te}
R.~Enberg, M.~H. Reno, and I.~Sarcevic
  \href{http://dx.doi.org/10.1103/PhysRevD.78.043005}{{\em Phys. Rev. D}
  {\bfseries 78} (2008) 043005}.

\bibitem{Gaisser:2012zz}
T.~K. Gaisser \href{http://dx.doi.org/10.1016/j.astropartphys.2012.02.010}{{\em
  Astropart. Phys.} {\bfseries 35} (2012) 801--806}.

\end{thebibliography}\endgroup

\clearpage
\section*{Full Author List: IceCube-Gen2 Collaboration}

\scriptsize
\noindent
R. Abbasi$^{17}$,
M. Ackermann$^{71}$,
J. Adams$^{22}$,
J. A. Aguilar$^{12}$,
M. Ahlers$^{26}$,
M. Ahrens$^{60}$,
C. Alispach$^{32}$,
P. Allison$^{24,\: 25}$,
A. A. Alves Jr.$^{35}$,
N. M. Amin$^{50}$
% ...
\\
\\
% ...
$^{12}$ Universit{\'e} Libre de Bruxelles, Science Faculty CP230, B-1050 Brussels, Belgium \\
% ...
$^{17}$ Department of Physics, Loyola University Chicago, Chicago, IL 60660, USA \\
% ...
$^{22}$ Dept. of Physics and Astronomy, University of Canterbury, Private Bag 4800, Christchurch, New Zealand \\
% ...
$^{24}$ Dept. of Astronomy, Ohio State University, Columbus, OH 43210, USA \\
$^{25}$ Dept. of Physics and Center for Cosmology and Astro-Particle Physics, Ohio State University, Columbus, OH 43210, USA \\
$^{26}$ Niels Bohr Institute, University of Copenhagen, DK-2100 Copenhagen, Denmark \\
% ...
$^{32}$ D{\'e}partement de physique nucl{\'e}aire et corpusculaire, Universit{\'e} de Gen{\`e}ve, CH-1211 Gen{\`e}ve, Switzerland \\
% ...
$^{35}$ Karlsruhe Institute of Technology, Institute for Astroparticle Physics, D-76021 Karlsruhe, Germany  \\
% ...
$^{50}$ Bartol Research Institute and Dept. of Physics and Astronomy, University of Delaware, Newark, DE 19716, USA \\
% ...
$^{60}$ Oskar Klein Centre and Dept. of Physics, Stockholm University, SE-10691 Stockholm, Sweden \\
% ...
$^{71}$ DESY, D-15738 Zeuthen, Germany \\
% ...

\subsection*{Acknowledgements}

\noindent
USA {\textendash} U.S. National Science Foundation-Office of Polar Programs,
U.S. National Science Foundation-Physics Division,
U.S. National Science Foundation-EPSCoR,
Wisconsin Alumni Research Foundation,
Center for High Throughput Computing (CHTC) at the University of Wisconsin{\textendash}Madison,
Open Science Grid (OSG),
Extreme Science and Engineering Discovery Environment (XSEDE),
Frontera computing project at the Texas Advanced Computing Center,
U.S. Department of Energy-National Energy Research Scientific Computing Center,
Particle astrophysics research computing center at the University of Maryland,
Institute for Cyber-Enabled Research at Michigan State University,
and Astroparticle physics computational facility at Marquette University;
Belgium {\textendash} Funds for Scientific Research (FRS-FNRS and FWO),
FWO Odysseus and Big Science programmes,
and Belgian Federal Science Policy Office (Belspo);
Germany {\textendash} Bundesministerium f{\"u}r Bildung und Forschung (BMBF),
Deutsche Forschungsgemeinschaft (DFG),
Helmholtz Alliance for Astroparticle Physics (HAP),
Initiative and Networking Fund of the Helmholtz Association,
Deutsches Elektronen Synchrotron (DESY),
and High Performance Computing cluster of the RWTH Aachen;
Sweden {\textendash} Swedish Research Council,
Swedish Polar Research Secretariat,
Swedish National Infrastructure for Computing (SNIC),
and Knut and Alice Wallenberg Foundation;
Australia {\textendash} Australian Research Council;
Canada {\textendash} Natural Sciences and Engineering Research Council of Canada,
Calcul Qu{\'e}bec, Compute Ontario, Canada Foundation for Innovation, WestGrid, and Compute Canada;
Denmark {\textendash} Villum Fonden and Carlsberg Foundation;
New Zealand {\textendash} Marsden Fund;
Japan {\textendash} Japan Society for Promotion of Science (JSPS)
and Institute for Global Prominent Research (IGPR) of Chiba University;
Korea {\textendash} National Research Foundation of Korea (NRF);
Switzerland {\textendash} Swiss National Science Foundation (SNSF);
United Kingdom {\textendash} Department of Physics, University of Oxford.

\end{document}